\documentclass[aps,preprint,tightenlines,superscriptaddress,
amsfonts,amssymb,amsmath,byrevtex,showpacs,nofootinbib,11pt]{revtex4-1}

\usepackage[english]{babel}
\usepackage[T1]{fontenc}
\usepackage{hyperref}
\hypersetup{colorlinks, allcolors=blue}

\newcommand{\be}{\begin{equation}}
\newcommand{\ee}{\end{equation}}

\begin{document}

\title{On extra dimensions and the cosmological constant problem}

\author{Grzegorz Plewa\footnote{Email: {\tt grzegorz.plewa@ncbj.gov.pl}}}
\affiliation{National Centre for Nuclear Research, Interdisciplinary Division for Energy Analyses, Wo\l{}odyjowskiego 83, 02-724, Warsaw, Poland}

\begin{abstract}
We consider a massive scalar field with a coordinate-dependent mass in higher-dimensional spacetime. The field satisfies Dirichlet boundary conditions on a brane representing the four-dimensional world. Despite being massive, the theory is scale-invariant. We quantize the theory calculating the zero-point energy. We find the lower bound for the uncertainty product in the uncertainty principle. We show that the zero-point energy density could be small if large extra dimensions exist. Identifying the zero-point energy as a source of dark energy, we extract the four-dimensional cosmological constant from higher-dimensional theory, considering quantum fluctuations close to the brane surface. We examine numerically ten- and eleven-dimensional spaces. The resulting zero-point energy is parameterized by the number of extra dimensions and the additional dimensionless {\it saturation parameter}, expressing the deviation from perfect saturation of the uncertainty principle. Letting the parameter to be small and of order of the fine-structure constant, we reproduce the experimental value of the cosmological constant in four dimensions.
\end{abstract}
\maketitle

\tableofcontents

\section{Intoduction}
\label{sec:intro}
The ADD hypothesis of large extra dimensions \cite{Arkani-Hamed:1998jmv, Randall:1999vf, Maartens:2010ar} provides an interesting alternative to the standard compactification. In a broader sense the concept is consistent with the idea that the four-dimensional universe can be just a hypersurface in a higher-dimensional space. This is predicted by the brane cosmology \cite{Brax:2004xh, Maartens:2010ar}. As shown in \cite{Arkani-Hamed:1998jmv}, large extra dimensions may solve the hierarchy problem. Assuming that only gravity can propagate in  higher-dimensional space explains why it is so weak. It has also been suggested that ADD  can shed some light on the cosmological constant problem \cite{Weinberg:1988cp}, providing the mechanism of decreasing the zero-point energy \cite{Arkani-Hamed:2000hpr, BURGESS2004383}. Although it is difficult to find a satisfactory explanation of the famous discrepancy of one hundred and twenty orders of magnitude \cite{Weinberg:1988cp}, the ADD concept may still be a useful option. 

Extra dimensions are essential for the string theory. Starting with the simplest bosonic string, the required number of dimensions is greater than four. The same holds for superstring theories and M-theory. However, string theory is not just about strings. It contains  D-branes, slolitonic objects where open string can end. The presence of D-branes introduces a non-perturbative region, revealing interesting dualities, like the  ADS/CFT correspondence \cite{Maldacena:1997re, Becker:2006dvp, Schwarz:1998fd, Aharony:1999ti, Hubeny:2014bla}. The AdS/CFT provides powerful tools for high energy physics. For instance, relativistic  hydrodynamics of a hot quark-gluon plasma created in ion colliders can be extracted from gravity \cite{Casalderrey-Solana:2011dxg}. This opens the doors for experimental, indirect application of higher-dimensional physics.  Despite the duality bases on the bulk-boundary correspondence, the higher-dimensional bulk need not to be interpreted as a real place in the universe. On the other hand, placing the universe on a brane is more than a mathematical correspondence and could have interesting consequences. For instance, the corresponding  gravitational effects could explain unresolved phenomena, like the nature of dark matter \cite{Abdusattar:2015azp}.

The standard explanation why we do not observe extra dimensions predicted by the string theory is that they are small and compact, too small to be detected.  From ADD point of view the answer could be different. Extra dimensions need not to be small. They are hidden because of the confinement on the brane. In particular, if the whole universe is just a D3-brane or another membrane of some kind, the internal observer would see no extra dimensions or they can be extremely difficult to observe. Still, the brane is a part of a multidimensional world and, as such, can interact with strings and different branes.

The zero-point energy is a natural source of vacuum energy density related with the cosmological constant. The problem is the huge discrepancy between cosmological energy density and theoretical predictions of quantum field theory \cite{Weinberg:1988cp}. In this paper we  address the issue utilizing basic elements of ADD hypothesis within the formalism of a scalar field theory. We will identify the universe to be a four-dimensional hypersurface, specified by Dirichlet boundary conditions for a massive scalar field in extra dimensions. The surface will represent the brane. In the simplified model we ignore the details of the brane action and its backreaction to the geometry. Instead, as in \cite{Arkani-Hamed:1998jmv}, we focus on some very basic aspects of extra dimensions. We allow the observer on the brane probing small distances in extra dimensions, however, only up to some small scale of order of the Planck length. This will express the confinement: it is possible to go beyond the four-dimensional bound but not very far.  From the observer's perspective extra dimensions are small. In reality, they can be compactified on a much bigger scale or not compact at all.

Mathematically the mechanism will be implemented considering a scalar field with coordinate-dependent mass term. The latter will be built out of momenta components needed to get insight into higher dimensions. The smaller the scale to probe, the more massive the field. Staying on a brane it is impossible to go beyond a given length scale and observe directly large extra dimensions.

We quantize the theory finding the zero-point energy density and providing a non-standard justification for the uncertainty principle. We use the resulting formula to construct four-dimensional cosmological constant. The latter will be parameterized by additional, dimensionless {\it saturation parameter}. It expresses the deviation from perfect saturation of the uncertainty principle. We observe that the cosmological constant and the corresponding zero-point energy density decrease exponentially with the saturation parameter and the number of extra dimensions. We show that it is possible to make them small even if considering Planck cut-off for momenta. We discuss models in ten and eleven dimensions. In both cases we reproduce the experimental value of the cosmological constant, finding the saturation parameter to be of order of the fine-structure constant.

\section{Scalar field with coordinate-dependent mass term}
\label{sec:classic}
Consider a $D$-dimensional Minkowski space and a four-dimensional brane\footnote{In this paper we reserve the term {\it brane} to a four-dimensional hypersurface in higher-dimensional space specified by Dirichlet boundary conditions for a scalar field.  We will ignore the fact that a brane, or more specifically, a D-brane, is defined by Dirichlet boundary conditions for open strings. Similarly, in this simplified approach we will not discuss the corresponding gravitational effects.} representing the four-dimensional world. Let $d$ stands for the number of orthogonal directions to the brane surface, so the number of extra dimensions. Let $X^A$, $A=0,...,D-1$ are position components in extra dimensions such that the first four components $x^{\mu} = (t, x^i)$ are coordinates on a brane while the rest, $\Delta x^j$, $j=1,...,d$, are extra dimensions. In particular, $X^A = (x^\mu, \tilde{x}^j) = (t, x^i, \tilde{x}^j)$ and $X^0 = x^0 = t$, $X^i = x^i$, $i=1,2,3$, $X^{j+3} = \tilde{x}^j$, $j=1,...,d$. Similarly, $P^A=(p^\mu, \tilde{p}^j)$ will stand for the momentum in extra dimensions. 

Suppose that a brane is just a rigid four-dimensional hypersurface, specified by the condition $\tilde{x}^j = 0$. We intensionally ignore details of the standard description of a brane, restricting to a scalar field in extra dimensions $\varphi(X)$. We require
\be
\label{base_dirichlet}
\varphi|_{\tilde{x}^j = 0} = 0,
\ee
i.e. the position of the brane is determined by the field. Imagine an observer on the brane that probes extra dimensions using high-energy particles.  Probing a length scale $\Delta \tilde{x}^j$ from $x^j=0$ to some non-zero position in the extra dimension requires a momentum of order of $(\Delta \tilde{x}^j)^{-1}$. It is a consequence of the uncertainty principle. Introducing a dimensionless parameter $a^i \geq \frac{1}{2}$ and $i =1,...,D-1$, the principle can be rewritten as an equation:
\be
\label{uncertainty_main}
\Delta X^{i} \Delta P^{i} = a^{i}.
\ee
The observer is confined on the brane in the sense that it is impossible to probe large extra dimensions. The confinement means that only a small scale in extra dimensions is accessible, possibly of order of Planck or string length. Using eq. \eqref{uncertainty_main} and utilizing the fact that the observer is localized on the surface $\tilde{x}^j = 0$, the required momentum to see the extra dimension reads
\be
\label{probing_momenta}
\tilde{p}^{j} \simeq \frac{a^{j}}{\Delta \tilde{x}^{j}} \simeq  \frac{a^{j}}{\tilde{x}^{j}}.
\ee
The last equality comes from the fact that the brane is localized in $\tilde{x}^{j} = 0$. In what follows, trying to probe the extra dimension from  $\tilde{x}^{j} = 0$ to some $\tilde{x}^{j} \neq 0$ means that the corresponding position uncertainty $\Delta \tilde{x}^{j}$ of a given test particle is of order of $\tilde{x}^{j}$\footnote{The position uncertainty {\it is not} the same as the distance scale, here represented by the difference of the two coordinates $x^{j} \neq 0$ and $x^j = 0$. However, being able to find a small  position uncertainty $\Delta x^j$ in extra dimension for a given particle, we are able to probe the corresponding small distance. The latter is of order of the position uncertainty.}, while the required momentum is of order of the corresponding momentum uncertainty. Here we are using the uncertainty principle to get the characteristic momentum associated to a given small distance scale. It is the momentum needed to probe the corresponding extra dimension.

The smaller  $\tilde{x}^{j}$, the more energy is needed. Let us remember that the brane is located at $\tilde{x}^{j} = 0$ and so $\tilde{x}^{j}$ refers to a relative distance scale as well (from perspective of the observer on the brane). As the relative distance, the latter can be associated with a  position uncertainty $\Delta \tilde{x}^{j}$ of a test particle. Alternatively, a virtual particle created spontaneously in the vacuum in this scale has a momentum of order of $a^j / \Delta \tilde{x}^{j}$ (assuming the particle exists); the same as the momentum of a test particle required to uncover the extra dimension.

The extra dimensions are not necessarily compact, but the observer is restricted to $| \tilde{x}^j| \leq \delta l$ because of the confinement. The latter is measured by $\delta l$ parameter. The perfect confinement means $\delta l = 0$ i.e. there is nothing to observe in extra dimensions from the brane perspective. In this case eq. \eqref{probing_momenta} gives infinite value: infinite energy is required to probe extra dimensions. On the other hand, if  $\delta l \neq 0$ the observer can probe extra dimensions, however, if $\delta l$ is small, a large momentum is required. In consequence, extra dimensions can be hidden even if they are not small. For the observer on the brane this additionally requires identifying another bound, $\varphi(\tilde{x}^i=\delta l)=0$. The condition reflects the boundary of the region $|\tilde{x}^i| \leq \delta l$. On the other hand, the field extends arbitrarily far in extra dimensions. Constructing the classical theory we temporarily ignore the condition  $|\tilde{x}^i| \leq \delta l$, leaving it as a subject for later analysis.

Consider the relativistic mass shell formula $E^2 = m^2 + {\vec{p} \,}^2$. Using eq. \eqref{probing_momenta}, the relation reads
\be
E^2 = m^2 + p_i p^i + M^2(X),
\ee
where
\be
\label{mass_term}
 M^2(X) = \sum_{i=1}^d  \frac{{a^i}^2}{{(\tilde{x}^i)}^2}.
 \ee
The term \eqref{mass_term} is the contribution to the energy from momenta in extra dimensions. Note that for the observer on a brane it can be interpreted as an additional contribution to the mass, labeled by coordinates $\tilde{x}^i$. If $m=0$ then it can be viewed as a coordinate-dependent mass. The latter expresses the fact that probing  extra dimensions is costly and requires more energy. Keeping this in mind consider the following action:
\begin{align}
  \label{action}
  S = \frac{1}{2} \int d^{D-1} X \Big(&\partial_A \varphi(X) \partial^A \varphi(X) -  M^2(X) \varphi^2(X) \Big),
\end{align}
where $M^2(X)$ is given by eq. \eqref{mass_term}. Note that it is the only non-zero contribution to the mass of the scalar field. As mentioned, the field extends in large extra dimensions, however, the observer on the brane is restricted to the region $|\tilde{x}^i| \leq \delta l$. The closer the brane, the greater the mass. Alternatively, starting with $\tilde{x}^i=0$ and probing extra dimensions on a scale $\tilde{x}^i \neq 0$ requires the momentum of order of $(\tilde{x}^i)^{-1}$.

Despite containing non-zero mass term, the action \eqref{action}  is scale-invariant. The symmetry would be broken by adding a constant mass. As discussed in \cite{Wetterich:2020cxq}, scale-invariance may be essential for fundamental theories. Following this intuition we will not consider a generalization involving non-zero constant mass. In fact, the symmetry may be a starting point to get the action \eqref{action}. More specifically, one can construct it searching for a coordinate-dependent term proportional to $\varphi^2$, such that the resulting scalar field theory is scale-invariant. This approach is much more elegant, but the meaning of the $a^i$ parameters is unclear. As we shall see in a moment, the connection with the uncertainty principle is reflected by solutions to classical equations of motions and the lower bound $a^{i}=\frac{1}{2}$ emerges as a separation between different classes of solutions.

Note that to get the  mass term \eqref{mass_term} we used the uncertainty principle to translate distance scales into the corresponding momenta. Technically the problem is that scales and position uncertainties are positive, while the coordinates  $\tilde{x}^j$ can be negative. We can keep the one-to-one correspondence between those things imposing the following symmetry for the solutions
\be
\label{symmetry}
\tilde{x}^{i} \rightarrow -\tilde{x}^{i}.
\ee
This symmetry is already reflected by the action \eqref{action}. The only thing is that we construct the solutions treating eq. \eqref{symmetry} as an extra requirement. We start with the following ansatz
\begin{align}
  \label{ans}
  &\varphi(X) = \psi(x) \Phi(\tilde{x}),
  \\[1ex]
   \label{facto}
  &\Phi(\tilde{x}) = \prod_{i = 1}^d \phi^{i}(\tilde{x}^{i}).
\end{align}
The factorization \eqref{ans} reflects the presence of the brane. Ignoring backreaction to the geometry it is reasonable to search for separable solutions, describing independently the hypersurface and extra dimensions.  The further factorization of the extra-dimensional part $\Phi$ is dictated by the uncertainty principle. For each dimension we have a simple, one-dimensional correlation between $\Delta x^{i}$ and the corresponding momentum $\Delta p^{i}$, reflected by the mass term \eqref{mass_term}.

The variation of the action \eqref{action} leads to the following equations of motion:
\begin{align}
  \label{wave}
  \partial_\mu \partial^\mu \psi(x) = 0,
  \\
  \label{phi_eq_oryg}
  \left(\frac{\partial^2}{\partial (\tilde{x}^{i})^2} + \frac{{a^i}^2}{(\tilde{x}^{i})^2}  \right)  \phi^{i}(\tilde{x}^{i}) = 0,
\end{align}
$\mu = 0,...,3$. The first equation, the wave equation \eqref{wave}, is the standard massless Klein-Gordon equation. The solution takes the standard form of the Fourier transform \cite{Srednicki:2007qs}:
\begin{align}
  \label{psi_sol}
  \psi(x) = \frac{1}{(2 \pi)^3 2 \omega} \int d^3 k \Big(a({\bf k}) e^{i k x} + a^*({\bf k})e^{- i k x} \Big),
  \\[1ex]
  \label{omega_form}
  \omega = \sqrt{k_1^2 + k_2^2 + k_3^3}.
\end{align}
Equation \eqref{phi_eq_oryg} can be simplified skipping labels of extra dimensions and writing in short:
\be
\label{phi_simple}
\phi''(\tilde{x}) + \frac{a^2}{\tilde{x}^2} \phi(\tilde{x}) = 0.
\ee
It is because we have the same form for each label $i$. Eq. \eqref{phi_simple} has solutions of three types. For $a > \frac{1}{2}$ one gets oscillating solutions of the form
\be
\label{osc_sol}
\phi(\tilde{x}) = A \sqrt{\frac{|\tilde{x}|}{l_0}} \cos\left( \frac{v}{2} \ln\frac{|\tilde{x}|}{l_0} \right) + B  \sqrt{\frac{|\tilde{x}|}{l_0}} \sin\left( \frac{v}{2} \ln\frac{|\tilde{x}|}{l_0} \right),
\ee
where
\be
\label{v}
v := \sqrt{4 a^2 -1}
\ee
and $A$, $B$ are integration constants and $l_0$ stands for dimensionfull parameter, used to make arguments of the logarithmic function dimensionless and ensuring correct normalization of the field (such that $A$, $B$ are dimensionless). We have only two integration constants and $l_0$ plays the role of a fixed normalization parameter. The absolute values were introduced to enforce the symmetry \eqref{symmetry}. In consequence, derivatives $\phi'(0)$, $\phi''(0)$ are ill defined at zero. However, it is not an issue since the mass term \eqref{mass_term} is already not well defined here. Divergent derivatives reflect the divergent  mass term.

The solution \eqref{osc_sol} is not scale-invariant in the standard sense $\tilde{x} \rightarrow \lambda  \tilde{x}$, but is scale invariant under the following, discrete transformations:
\begin{align}
  \label{discrete scalling}
  \tilde{x} \rightarrow \frac{l_n}{l_0} \tilde{x},
\end{align}
where
\be
\label{ln}
l_n = l_0 \exp \left( \frac{4 \pi n}{v} \right), \quad  n \in \mathbb{Z}.
\ee
It is straightforward to see that the action \eqref{action} transforms homogeneously under any of such transformations.  Note that eq. \eqref{discrete scalling} is  governed by the uncertainty parameter $a$. The closer to saturation, the higher the discrepancy between different $l_n$.

In case $a = \frac{1}{2}$ one finds the following solutions: 
\be
\label{one_half}
\phi = A \sqrt{\frac{|\tilde{x}|}{l_0}} + B \sqrt{\frac{|\tilde{x}|}{l_0}} \ln \left( \frac{|\tilde{x}|}{l_0} \right).
\ee
They are scale invariant if $B=0$.

Finally, the case $a < \frac{1}{2}$ corresponds to non-oscillating solutions of the form:
\be
\label{non_osc}
\phi = \sqrt{\frac{|\tilde{x}|}{l_0}} \left(A \left(\frac{|\tilde{x}|}{l_0}\right)^{- u/2} +  B  \left(\frac{|\tilde{x}|}{l_0}\right)^{u/2}\right),
\ee
where
\be
u := \sqrt{1 - 4a^2},
\ee
and $A$, $B$ are integration constants. It is straightforward to observe that the solution \eqref{non_osc} is scale invariant only if $A=0$ or $B=0$. On the other hand $a < \frac{1}{2}$ is inconsistent with the uncertainty principle. Therefore we will not discuss the form \eqref{non_osc}. As we shall see, there are additional restrictions at quantum level, making the class of solutions even smaller and consistent with the uncertainty principle.

Quite interestingly, we found three different forms of solutions to the simple equation \eqref{phi_simple} and the separation is given by the boundary value of $a$ parameter (in the sense of the uncertainty principle). Each of those classes shares different transformation properties. Requiring the uncertainty principle to be satisfied eliminates the solution \eqref{non_osc}.

\section{Quantization}
\label{sec:quant}
We now quantize the field $\varphi(X) = \psi(x) \Phi(\tilde{x})$, where $\Phi =  \prod_{i=1}^d\phi(\tilde{x}^{i})$ (see eq. \eqref{facto}) and $\phi$ are given by classical solutions consistent with the uncertainty principle, i.e.  eqs. \eqref{osc_sol} or \eqref{one_half}. Consider the following Fourier transform of the extra-dimensional part
\be
\label{b_form}
\Phi(\tilde{x}) = \frac{1}{(2 \pi)^d \, l_0^{d/2}} \int d^d \tilde{k} \left( b(\tilde{k})e^{i \tilde{k} \tilde{x}} + b^*(\tilde{k}) e^{-i \tilde{k} \tilde{x}}  \right).
\ee
The coefficient $l_0^{d/2}$ (from the original form of $\phi$) is to ensure correct normalization of the resulting Hamiltonian. In what follows
\be
\label{product_transforms}
\varphi(X) = \frac{l_0^{-d/2}}{(2 \pi)^{d+3} 2 \omega}  \int d^3 k \, d^d \tilde{k} \left( a({\bf k}) e^{i k x} + a^*({\bf k})e^{-i k x} \right) \left( b(\tilde{k})e^{i \tilde{k} \tilde{x}} + b^*(\tilde{k}) e^{- i \tilde{k} \tilde{x}}  \right),
\ee
where $\omega$ is given by eq. \eqref{omega_form}. Defining: $q_A = \{k_0, k_1,...,k_3,\tilde{k}_1,...,\tilde{k}_d \}$, one rewrites eq. \eqref{product_transforms} as a single transform:
\be
\label{simple_transform}
\varphi(X) =  \frac{1}{(2 \pi)^{d+3} 2 \omega  \, l_0^{d/2}}  \int d^d q \left( \alpha({\bf q}) e^{i q X} +  \alpha^*({\bf q})e^{-i q X} \right),
\ee
where
\be
\alpha(q) =  l_0^{d/2} \int d^{D-1} X e^{- i q X} \left( i \partial_0 \varphi + \omega \varphi \right).
\ee
The conjugate momenta are $\Pi(X) = \dot{\varphi}(X)$, while the Hamiltonian density takes the form:
\be
\label{hamiltonian_density}
{\cal{H}} = \frac{1}{2} \Pi^2 + \frac{1}{2} \partial_i \varphi \partial^i \varphi + \frac{1}{2} \partial_{\, \tilde{j}} \varphi \partial^{\, \tilde{j}} \varphi +
\frac{1}{2} \Delta m^2 \varphi^2,
\ee
$\partial_i = \partial / \partial x^i$, $\partial_{\, \tilde{j}} := \partial / \partial {\tilde{x}}^j$. The standard quantization conditions for $\varphi(t, {\bf X})$ and $\Pi(t, {\bf X})$ translate into the following relations for creation and annihilation operators:
\begin{align}
  \nonumber
  &[\alpha({\bf q}), \alpha({\bf q'})] = [\alpha^\dag({\bf q}), \alpha^\dag({\bf q'})] = 0,
  \\[1ex]
   &[\alpha({\bf q}), \alpha^\dag({\bf q'})] = (2 \pi)^{d+3} 2 \omega  \, l_0^{d/2} \delta^{d+3}({\bf q} - {\bf q'}).
\end{align}
This leads to the following vacuum expectation value of the Hamiltonian:
\begin{align}
  \label{H0}
  &\langle H \rangle_{vac} = H_{0}^{std} +  H_{0}^{ext},
  \\[1ex]
  \label{H_std}
  &H_{0}^{std} = \frac{1}{2} \frac{\delta^{d+3}(0)}{(2 \pi)^{d+3} l_0^d} \int d^3 k \, \omega,
  \\[1ex]
  \label{H_ext}
  &H_{0}^{ext} = \frac{d}{4} \frac{\delta^{d+3}(0)}{(2 \pi)^{d+3} l_0^d} \int d^3 k \, \omega^{-1} \int d \tilde{k} \, \tilde{k}^2.
\end{align}
Following the convention of skipping extra dimensional labels, in the last term of eq. \eqref{H_ext} $\tilde{k}$ stands for any of the momentum components $\tilde{k}_i$. We have $d$ identical terms and so $d$ identical integrals. Eq.  \eqref{H_std} is the standard zero-point energy \cite{Srednicki:2007qs}, reducing to the standard formula in the limit $d \rightarrow 0$. The term \eqref{H_ext} is the extra correction to the vacuum energy that comes from extra dimensions and the coordinate-dependent mass. It contains a divergent integral $\int d^3 k \omega^{-1}$, which goes to infinity when the frequency goes to zero. This infrared divergence is multiplied by an ultraviolet divergence, built out of the momenta $\tilde{k}$:
\be
H_{0}^{ext} \propto  \tilde{K}, \quad \tilde{K} := \int d \tilde{k} \,  \tilde{k}^2.
\ee
The latter can be easily regularized assuming that the   the function $\phi$ field satisfies Dirichlet boundary conditions at some large but finite $L \gg \delta l$:
\be
\label{Lperiod}
\phi |_{\tilde{x} = 0} = \phi |_{\tilde{x} = L} = 0.
\ee
Again, we have the same relation for each extra dimension $\tilde{x}^i$. The condition \eqref{Lperiod} is consistent with the oscillating solution \eqref{osc_sol} where $A=0$. For the Fourier transform \eqref{b_form} it means that the momenta $\tilde{k}$ are quantized: 
\be
\label{kn}
\tilde{k}_n = \frac{n \pi}{L}.
\ee
We will keep the notation convention  writing $\tilde{k}_n$ in short (instead of $(\tilde{k}_{i})_n$). In what follows:
\be
\tilde{K} \propto \sum_n \frac{n^2 \pi^2}{L^2} \propto  \sum_l l^2.
\ee
Using analytical continuation of the Riemann zeta function, one finds
\be
\tilde{K} \propto \zeta(-2) = 0.
\ee
The zero above can be interpreted as representing the difference between finite configuration of fixed $L$ and $L=\infty$:
\be
\label{rel}
\tilde{K}|_{L = \infty} - \tilde{K}|_{L = finite} = \frac{\pi^2}{L^2} \left(\int l^2 dl - \sum_l l^2 \right) = 0.
\ee
This results from direct application of the Euler-Maclaurin formula. Note that both infrared and ultraviolet divergences can be eliminated restricting to oscillating solutions \eqref{osc_sol} and interpreting  $\Delta K$ term as representing relative contribution to the energy. In what follows, $H_0^{ext}=0$, and $H_0^{std}$ is the only non-zero term in the zero-point energy \eqref{H0}.

\section{Uncertainty principle}
Constructing the action we observed that already at the classical level the scalar field reproduces the lower bound $a=\frac{1}{2}$ in the uncertainty principle. We obtained three groups of solutions: $a>\frac{1}{2}$, $a=\frac{1}{2}$ and $a < \frac{1}{2}$. The latter is inconsistent with the uncertainty principle, however, cannot be ruled away at the classical level.

Quantizing the field we noticed the presence of the peculiar term $H_0^{ext}$ in the zero-point energy, which is both IR and UV divergent. We showed that this term can be effectively eliminated imposing Dirichlet boundary conditions \eqref{Lperiod}. This can be done restricting  to the oscillating solutions built from only the sine term. The are present for $a>\frac{1}{2}$.

The logic can be reversed and one can use the result to justify the uncertainty principle. The principle can be viewed as something which improves the zero-point energy. Indeed, the elimination of the peculiar term $\tilde{K}$ is only possible if restricting to oscillating solutions \eqref{osc_sol}, and so $a > \frac{1}{2}$. In what follows:
\be
\label{unc_final}
\Delta x \Delta p > \frac{1}{2}.
\ee
Note that the boundary value $a = \frac{1}{2}$ has been excluded. In general, it does not mean that $a$ is always greater than one-half, preventing perfectly Gaussian wave packages to exist. Still, $a$ can be arbitrarily close to one-half. From experimental perspective the lack of the single value $a=\frac{1}{2}$ is meaningless.

To get the result we used the standard form of the uncertainty principle based on the product of position and momentum uncertainties. However, the presence of minimal length scale in theories of quantum gravity suggests a modification to the standard position-momentum commutation relation \cite{Kempf:1994su, Bosso:2020aqm}):
\be
\label{xp_mod}
[\hat{x}, \hat{p}] = i f(\hat{p}).
\ee
For the uncertainty principle it means that the presence of  terms square in momentum uncertainty $\Delta p^2$ and others is expected \cite{Bosso:2020aqm}. Clearly, the modification would change a lot in the model we discussed, since the form of coordinate-dependent mass reflects the form of the uncertainty principle. Modifying the term makes the theory much  more complicated and we will not consider this in the paper. However, there is one thing that is worth mentioning. Constructing the classical action we noticed the presence of scale-invariance. As discussed in \cite{Wetterich:2020cxq}, this symmetry may be essential constructing a fundamental theory. In fact, it was the main reason why we did not incorporate a constant non-zero mass.  Now, modifying the uncertainty principle modifies the coordinate-dependent mass, breaking the symmetry. In what follows, the modification seems to be at least technically problematic. On the other hand, it would be desirable to get a lower bound $a > \frac{1}{2}$. As we shall see in the next section, the presence of such bound, slightly greater than one-half, could explain the presence of small, non-zero cosmological constant.

\section{Simple ADD model and the zero-point energy density}

We now use the quantization results to get the four-dimensional cosmological constant. Doing so we will identify the zero-point energy of the scalar field as a candidate for dark energy. 

Consider an observer on the hypersurface $\tilde{x}^i = 0$. We will continue to use the simplified notation in which we skip the extra dimensional labels. So, if $D=10$, like in case of superstring theories, there are six orthogonal directions to the hypersurface. As previously, we associate the surface with the brane ignoring the details of its gravitational footprint. 

Note that there are two dimensional length scales: the probing scale $\delta l$, expressing the effective size of extra dimensions from a brane perspective and a scale given by the auxiliary regularization parameter $L$. For the first time we incorporate the confinement on the brane, requiring:
\be
\label{confi}
\varphi|_{\tilde{x} = \delta l} = 0.
\ee
Eq. \eqref{confi} should be supplemented by previous boundary conditions \eqref{Lperiod}. All they are satisfied by oscillating solutions \eqref{osc_sol} if $l_0 = \delta l$ or $l_0 = L$. We adopt the latter choice\footnote{With this choice the $L$ parameter in eq. \eqref{kn} is played by $\delta l$ while the function $\phi$ vanishes identically in $\tilde{x} = L$. However, this does not affect regularization procedure presented in sec. \ref{sec:quant}. It is just discretization on a different scale.} because then $\varphi \rightarrow 0$ as $L \rightarrow  \infty$. The corresponding oscillating function $\phi$ reads:
\be
\label{osc_sol_final}
\phi(\tilde{x}) = B  \sqrt{\frac{|\tilde{x}|}{L}} \sin\left( \frac{v}{2} \ln\frac{|\tilde{x}|}{L} \right).
\ee
The boundary conditions \eqref{Lperiod} and \eqref{confi} imply
\be
\label{Ln}
\frac{\delta l}{L} = \exp \left( - \frac{2 \pi n}{v_n} \right), \quad n=1,2,3...
\ee
or, alternatively
\be
\label{v_quant}
v_n = v_1 n, \quad v_1 = \frac{2 \pi}{\ln(L / \delta l)},
\ee
where $v_n = \sqrt{4 a_n^2 -1}$ represents the form built out of quantized $a_n$. The coefficient is discrete as a direct consequence of eqs. \eqref{Lperiod} and \eqref{confi}. Letting $L \gg \delta l$, the smallest $a_n$ is for $n=1$.

Let $\delta E_0$ be a total amount of zero-point energy for the observer on the brane. It is the part of the total energy $H_0^{std} L^{D-1} $, associated with a small region in extra dimensions close to the brane surface:
\be
\label{e_init}
\delta E_0 =  H_0^{std} \left( \frac{\delta l}{L} \right)^d.
\ee
Note that discussing the confinement we assumed that the observer on the brane has access only to a small region in extra dimensions $|\tilde{x}^i| \leq \delta l$ (the brane is placed at $\tilde{x}^i = 0$). In eq. \eqref{e_init} $\delta l^d / L^d$ determines the fraction of the total energy accessible for the observer as a simple quotient of the corresponding volumes. The four-dimensional energy density reads
\be
\delta \rho_0 = \frac{H_0^{std}}{L^3} \left( \frac{\delta l}{L} \right)^d.
\ee
It is a total zero-point energy close to the boundary (including the higher-dimensional part), divided by regularized three-dimensional volume $L^3$.
Recalling the form of the zero-point energy \eqref{H_std}, substituting $l_0 = L$ and using standard interpretation of the Dirac delta at zero, $\delta^{d+3}(0) = (2 \pi)^{d+3} L^{d+3}$, one finds
\be
\delta \rho_0 = \frac{1}{2}  \left( \frac{\delta l}{L} \right)^d  \int d^3 k \omega.
\ee
The integral above can be regularized using the standard ultraviolet cut-off for momenta $k_i$. Since $\delta l$ is the boundary in extra dimensions for the observer on the brane, it is expected that it is fundamental, being possibly of order of string or Planck scale. It translates into the corresponding cut-off scale for the momentum: $k_{cut} = a_n \delta l^{-1}$. Here we used the exact form of the uncertainty principle with quantized $a$ labeled by $n$ (see eq. \eqref{Ln} and the discussion below). Performing the integral leads to
\be
\label{n_dens}
\delta \rho_0 = \frac{9 \pi  a_n^4}{2 \, \delta l^4}  \left( \frac{\delta l}{L} \right)^d.
\ee
The form above is parameterized by $a_n$ coefficient. The greater $a_n$, the bigger the zero-point energy. Note that the Dirichlet boundary conditions do not allow the standard lower bound  $a=\frac{1}{2}$. Instead, for finite $\delta l$ and $L$, we get the minimum for some $a_1 > \frac{1}{2}$. The value is determined by the ratio of the length scales $\delta l$ and $L$. For $L \gg \delta l$, $a_1$ can be arbitrarily close to one-half. Using eq. \eqref{Ln} with  $a_n = a_1$, one rewrites  \eqref{n_dens} as
\be
\label{dens_min}
\delta \rho_{min}  = \frac{9 \pi  a_1^4}{2 \, \delta l^4}  \exp \left( - \frac{2 \pi d}{v_1} \right).
\ee
Since $a_1 \neq \frac{1}{2}$, the energy density \eqref{dens_min} is non-zero.

Deriving this form we skipped all labels of extra dimensions. In particular, in the adopted convention we were writing $a_1$ instead of $a_1^i$, $i=1,...,d$. The notation ignores the fact that we can have different $a$ per spacetime dimension. This assumption can be justified by homogeneouty and isotropy of space. Being interested in finding the minimum energy density, there is no reason to distinguish between different directions in space and, in particular, considering different $a_1$ per space-like direction. 

\section{The cosmological constant problem}
The zero-point energy density is a natural candidate for the cosmological constant. However, the standard derivation involving Planck cut-off for energy leads to enormous energy density and the cosmological constant  more than one hundred twenty orders of magnitude bigger than the observed one. It is the famous cosmological constant problem  \cite{Weinberg:1988cp}.

We now consider the energy density \eqref{dens_min} as a candidate for the vacuum energy related with the cosmological constant. We will make two assumptions: {\it i)} the small scale $\delta l$ is comparable with the Plank length $l_P$ and, in particular, we identify $\delta l = l_P$, {\it ii)} there exists a lower bound $a = a_1 > \frac{1}{2}$, specifying the minimum in eq. \eqref{dens_min}. The second assumption is nothing but the statement that the Heinsenberg uncertainty principle cannot be perfectly saturated, i.e. the $a$ parameter is always greater than one-half. In other words, there is a parameter determining the value of this minimum.  Note that the inability of saturating the principle is a consequence of the boundary conditions \eqref{Lperiod}-\eqref{confi}, making $a$ discrete. Also note that eq. \eqref{dens_min} comes from the higher-dimensional formula, in which we restricted to a region close to the brane surface. We divided the corresponding energy by $L^3$ to get the standard energy density in four dimensions. Now we will examine the consequences.

Letting $\delta l = l_P$, from eq. \eqref{dens_min} one finds the following cosmological constant:
\be
\label{cos}
\Lambda = \frac{36 \pi^2  a^4}{l_P^2}  \exp \left( - \frac{2 \pi d}{\sqrt{4 a^2 -1}} \right).
\ee
We skipped ``$1$'' label in $a$ understanding that we are talking about the smallest possible $a > \frac{1}{2}$. The cosmological constant \eqref{cos}, as well as the corresponding zero-point energy density, decreases with the number of extra dimensions $d$. It is similar to the gravitational leakage effect, explaining the gravitational weakness in ADD models. In contrast to gravity, however, the mechanism is less intuitive.  We should also keep in mind that eq. \eqref{cos} was derived in a higher-dimensional flat Minkowski space. In particular, we intensionally ignored  gravitational effects caused by the presence of a brane. 

The expression \eqref{cos} can be small. According to \cite{Planck:2018vyg}, the experimental value of the cosmological constant is  $\Lambda_{obs} \simeq 2.8 \times 10^{-122}$   (in Planck units). Let $\Delta a$ be a {\it saturation parameter}, defined by:
\be
a = \frac{1}{2} + \Delta a.
\ee
It is a small non-zero parameter, expressing the deviation from perfect saturation of the uncertainty principle. The smaller $\Delta a$, the smaller the cosmological constant \eqref{cos}. Letting $d=6$, so as for superstring theories, one finds $\Lambda  \simeq 2.29 \times 10^{-122}$ for $\Delta a \simeq 0.00441$. In case of $D=11$ ($d=7$), like for M-theory or eleven-dimensional supergravity, one gets $\Lambda  \simeq 2.74 \times 10^{-122}$ for $\Delta a \simeq 0.006$. Quite interestingly, in both cases extremely small value of the cosmological constant was obtained taking $\Delta a$ of order of the fine-structure constant $\alpha \simeq 0,0072$. This shows that the presence of small, non-zero cosmological constant can be understood as a consequence of zero-point energy of the scalar field in extra dimensions. Letting $d=0$, i.e. assuming that there is no extra dimensions, eq. \eqref{cos} leads to large cosmological constant and the cosmological constant problem. In what follows, the simple formula \eqref{cos} provides a mechanism of how the problem can be solved. However, ignoring the details of the brane physics, it is not a complete solution of the problem.

An open question is the meaning of the lower bound for the saturation parameter. This can be a result of topological obstructions in space, like the presence of another brane. In fact, fixing the two scale parameters: $\delta l$ and $L$, one gets the lower bound for $a$. Promoting $L$ to the real parameter could explain the origin of the minimum.

Another option is modification of the uncertainty principle \cite{Kempf:1994su, Bosso:2020aqm}. This has been already discussed in the paper. We mentioned potential problems with complexity and breaking scale-invariance. Quantum gravitational effects may also affect the result, explaining the presence of non-zero bound for $\Delta a$. However, this is beyond the scope of the paper, requiring more realistic models and considering string theory effects.

Finally, looking for a reason why the saturation parameter is small but non-zero, one could consider even simpler explanation. A wave packet saturating perfectly the uncertainty principle (Gaussian or another one with $\Delta a = 0$), passing through a rectangular slit changes such that $\Delta a$ is non-zero \cite{Nairz:2002}. In what follows, simple obstructions in space could explain small deviation from the perfect Gaussian case and perfect saturation of the uncertainty principle. This has nothing to do with quantum gravity, higher dimensions and branes. It is just a consequence of the presence of objects in space. It would be interesting to ask if the lower bound of $a$ could be found in that way. On the other hand, a more subtle quantum effects could make $\Delta a$ non-zero, changing it in the same way as bare parameters in quantum field theory, e.g. anomalous magnetic momentum.

\section{Summary}
In this paper we considered a very simple model of a massive scalar field in extra dimensions. Within this model the observer on the brane could probe extra dimensions within a  small length scale, of order of the fundamental length. The mechanism was incorporated introducing a scalar field with coordinate-dependent mass term. Its form was dictated by the uncertainty principle, rewritten as an equation with a dimensionless $a$ parameter. We observed that despite being massive, the scalar field theory is scale-invariant. 

Depending on the value of the parameter we identified three classes of classical solutions. For $a > \frac{1}{2}$ we get oscillation solutions, the boundary value $a = \frac{1}{2}$ corresponds to non-oscillating, partially logarithmic functions, while for $a < \frac{1}{2}$ we obtained hyperbolic non-oscillating solutions.
Only the first two classes are consistent with the uncertainty principle. The exact scale-invariance is present only for the second and the third class, and only for certain values of  parameters. The oscillating solutions break the standard continuous scale-invariance, however, a discrete version of this symmetry remains. The boundary value $a=\frac{1}{2}$ emerges already at the classical level. The classical theory cannot justify the uncertainty principle, since the non-oscillating solutions for $a < \frac{1}{2}$ cannot be ruled away at the classical level.

We quantized the theory finding the zero-point energy. We observed that in addition to the standard UV-divergent term there is another one, that comes from extra dimensions. The term was even  more problematic, involving both ultraviolet and infrared divergences. We showed that both of them can be easily removed imposing Dirichlet boundary conditions at a large, auxiliary scale $L$, and using the standard Riemann zeta function or Euler-Maclaurin summation.

Based on the quantum result we proposed a simple model of the cosmological constant. In this model the vacuum energy density in four dimensions is given by the zero-point energy of the scalar field in  a rectangular region close to the brane surface. The size of the region in extra dimensions is specified by a small scale $\delta l$, which determines the maximum distance from perspective of the observer on the brane. We considered the Planck length as the value of this parameter, letting the cut-off for momenta to be specified by the inverse of this length scale. Assuming that $L$ is large but finite we showed that $a$ is quantized and the minimum value of the energy density in the vacuum  corresponds to the minimum  $a > \frac{1}{2}$. In the limit $a \rightarrow \frac{1}{2}$  the zero cosmological constant goes to zero. However, for finite $L$ the zero-point energy is always non-zero. The auxiliary scale $L$ was eventually eliminated from the result.

Choosing $a$ properly it is possible to reproduce the experimental value of the cosmological constant. The latter could be small even with the Planck cut-off for energy.  The reason is that the zero-point energy  decreases exponentially with the number of extra dimensions $d$ and the saturation $\Delta a$. For $d=6,7$, as predicted by superstring theories and M-theory, the experimental value $10^{-122}$ in Planck units corresponds to the saturation parameter  of order of the fine-structure constant. In what follows, there is no cosmological constant problem and the zero-point energy density is small.

We discussed various possible explanations of a non-zero lower bound for the saturation parameter $\Delta a$. This includes topological obstructions in higher-dimensional space, the presence of another brane and quantum gravity effects. We also mentioned a practical inability to construct and keep perfectly Gaussian wave packet, which corresponds to the zero saturation parameter. If treated seriously, this inability translates into the condition $\Delta a \neq 0$, leading to small but non-zero cosmological constant. Another explanation could be the presence of quantum corrections to $a$ parameter, similar to vacuum polarization. These corrections need not to be at the fundamental scale, since the required saturation parameter is close to the fine-structure constant.

Examining the results we provided a non-standard explanation of the uncertainty principle. The boundary value $a=\frac{1}{2}$ was found already at the classical level. Quantizing the theory  we get the condition $a > \frac{1}{2}$ as a constraint eliminating part of divergent terms from the zero-point energy.

We did not incorporate fully the physics of the brane in the model and, in particular, its backreaction to the geometry. Instead, we restricted to the scalar field with coordinate-dependent mass. The idea was to show that large extra dimensions can be useful solving the cosmological constant problem i.e. identifying the dark energy as a direct consequence of the zero-point energy. The final form of the cosmological constant in the model was parameterized by two scales: the Planck cut-off scale  and the saturation parameter. An interesting thing is the connection between the number of extra dimensions, $\Delta a$ and the zero-point energy in the vacuum. The presence of non-zero saturation parameter could explain the presence of non-zero cosmological constant. It is possible that measuring $\Delta a$ with high enough precision could tell something about the value of the fundamental scale and the cosmological constant, uncovering details of higher-dimensional physics. However, experimental verification of the Heisenberg uncertainty principle is not easy \cite{Nairz:2002, Schurmann:2022znj, Matteucci:2010}. It would be extremely difficult, if at all possible, to imagine an experiment estimating the lower bound of $\Delta a$. Still, results presented in this paper suggest a correlation between this bound and the value of the cosmological constant.

The model can be generalized in many different ways. For instance, one could consider a more detailed description of the brane, including string effects. In particular, it would be interesting to examine the consequences of incorporating the scalar field with coordinate-dependent mass. An open question is the potential meaning of the scale-invariance. As remember, the symmetry was a consequence of the uncertainty principle. Fixing the mass term makes the scalar field theory scale-invariant. However, the symmetry is not fully reflected by the solutions. What is more, any generalization of the uncertainty principle at the fundamental scale could break the symmetry right from the beginning, making the theory much more complicated. 

Finally, an intriguing question is the expected value of lower bound for the saturation parameter and the fact, it is so close to the fine-structure constant. This may suggest that processes far from the Planck scale may be crucial in understanding the microscopic nature of vacuum energy.

\end{document}